\documentclass[aps,pra,reprint,twocolumn,superscriptaddress,floatfix]{revtex4}

\usepackage{amsmath}
\usepackage{graphicx}
\usepackage{dcolumn}
\usepackage{mathrsfs}
\usepackage{multirow}
\usepackage{mciteplus}
\usepackage{color}
\usepackage{epstopdf}

\newcommand{\braket}[2]{\ensuremath{\left\langle #1 \vphantom{#2} \right| \left. #2 \vphantom{#1} \right\rangle}} 
\newcommand{\bracket}[3]{\ensuremath{\left\langle #1 \vphantom{#2}\vphantom{#3} 
           \right| #2 \left| #3 \vphantom{#1} \vphantom{#2} \right\rangle}} 
\newcommand{\bra}[1]{\ensuremath{\left\langle #1 \right|}}
\newcommand{\ket}[1]{\ensuremath{\left| #1 \right\rangle}}

\begin{document}


\title{Variational treatment of electron-polyatomic molecule scattering calculations using adaptive overset grids} 


\author{Loren Greenman}
\email{lgreenman@lbl.gov}
  \affiliation{Chemical Sciences Division, Lawrence Berkeley National Laboratory, Berkeley CA 94720}
  \affiliation{Department of Chemistry, University of California, Davis, CA 95616 USA}
\author{Robert R. Lucchese}
  \affiliation{Chemical Sciences Division, Lawrence Berkeley National Laboratory, Berkeley CA 94720}
  \affiliation{Department of Chemistry, Texas A\&M University, College Station, TX 77843 USA}
\author{C. William McCurdy}
  \affiliation{Chemical Sciences Division, Lawrence Berkeley National Laboratory, Berkeley CA 94720}
  \affiliation{Department of Chemistry, University of California, Davis, CA 95616 USA}
\date{\today}

\begin{abstract}
The Complex Kohn variational method for electron-polyatomic molecule scattering
is formulated using an overset grid representation of the scattering wave function. 
The overset grid consists of a central grid
and multiple dense, atom-centered subgrids that allow the simultaneous spherical 
expansions of the wave function about multiple centers.  Scattering boundary conditions
are enforced by using a basis formed by the repeated application of the  free particle
Green's function and potential, $\hat{G}^+_0\hat{V}$ on the overset grid in a ``Born-Arnoldi'' solution 
of the working equations.   The theory is 
shown to be equivalent to a specific Pad\'e approximant to the $T$-matrix, and has 
rapid convergence properties, both in the number of numerical basis functions employed and
the number of partial waves employed in the spherical expansions.  The method is demonstrated
in calculations on methane and CF$_4$ in the static-exchange approximation, and compared in 
detail with calculations performed with the numerical Schwinger variational approach based on 
single center expansions. An efficient procedure for operating with the free-particle Green's function and 
exchange operators (to which no approximation is made) is also described. 
\end{abstract}

\maketitle

\section{Introduction}
\label{sec:intro}

In recent decades, three \emph{ab initio} approaches to electron-molecule collisions and molecular photoionization have been developed and widely applied to polyatomic molecular targets:  (1) the Complex Kohn variational method~\cite{mccurdy1989collisions,lengsfield_methane_1991,parker_H2_close_coup_1991,lengsfield1991electronmolecule,rescigno1995incorporation,rescigno1995complex,trevisan_imaging_2012,douget_DEA_CH4_2015}, (2) the Schwinger variational method~\cite{lucchese1981iterative,lucchese1982studies} and (3) the R-matrix method~\cite{Tennyson_review_2010}.  The first two of these approaches are based explicitly on variational principles and their accuracy and applicability derives from the form of the trial functions employed.  The third, the R-matrix method, is an embedding approach based on dividing space into an inner region containing the molecule and an outer region in which the interactions are simpler.   These approaches make different compromises in their combined numerical treatments of electronic correlation, the bound electronic states of the target molecules or ions, and the solution of the highly nonspherical electron-molecule scattering problem itself.  Consequently all three have well-recognized limitations in the size of molecules that they can treat practically.  More importantly, the accuracy that can be expected of these methods when they are applied to molecules containing more than a few first row atoms is limited by practical restrictions on the quality of the treatment of correlation and target response for larger systems, including the number and accuracy of the target electronic states that can be included. 
\begin{figure}[h]
\begin{tabular}{c}
\includegraphics[width=0.95\columnwidth]{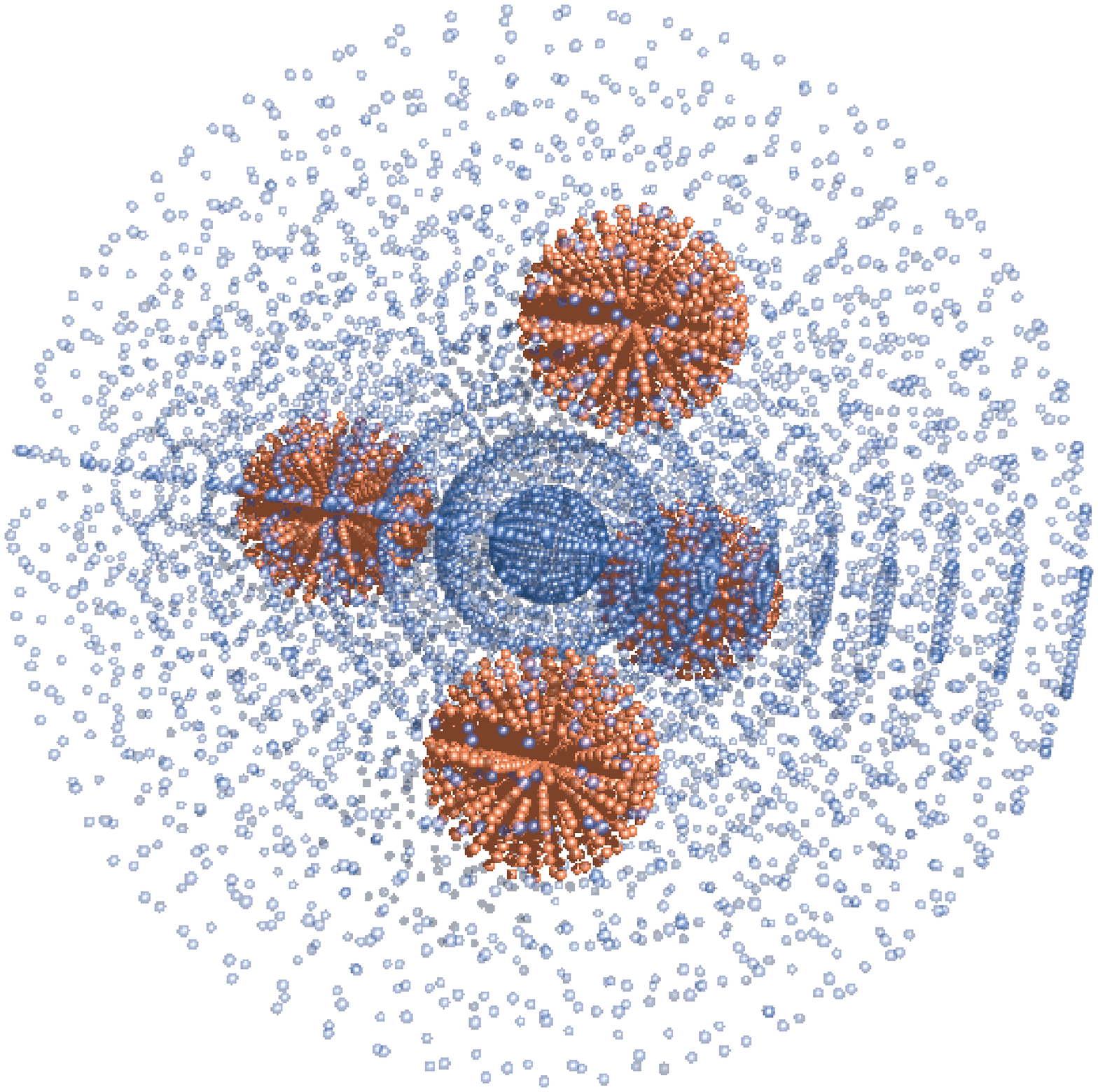} \\
\includegraphics[width=0.75\columnwidth]{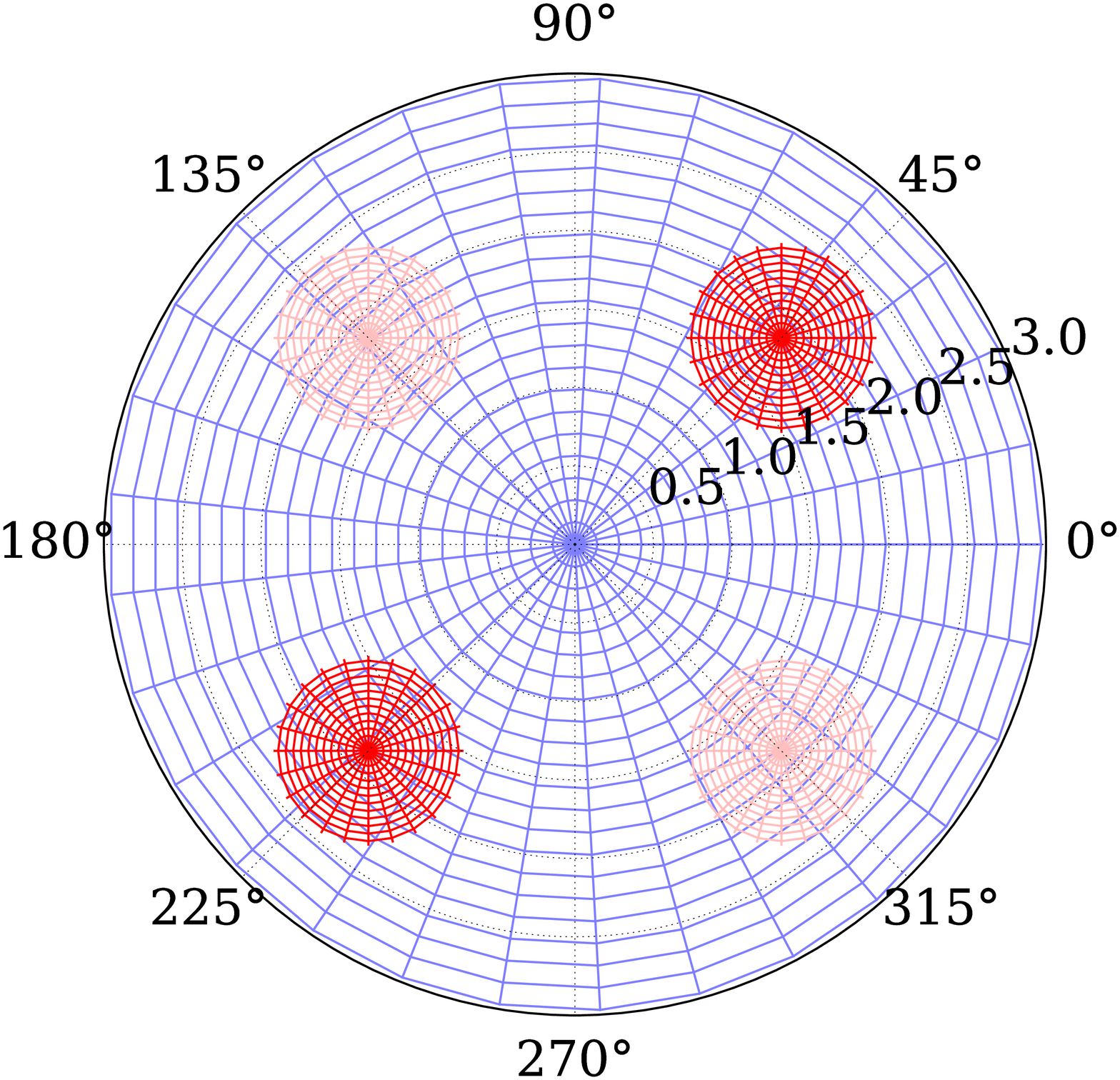}
\end{tabular}
\caption{The overset grids for CF$_4$, in 3-D (top panel) and in a
projection (bottom panel, darker shade is coming out of the page).
These are sketches of the overset grid  with relatively few points and do not represent
the much denser and adaptively spaced grids used in the calculations.
\label{fig:CF4OversetGrid}}
\end{figure}

Those limitations have their origin in the compromises made to combine the target electronic structure and correlation aspects of electron-molecule collisions with the specific method used to solve the scattering problem.   In this work we demonstrate a new version of the Complex Kohn approach that solves the scattering problem entirely using adaptive grid-based methods and that substantially extends its applicability to a wider range of physical problems, including processes involving core excited states at very high energies or diffuse Rydberg-like states of polyatomic molecules.   It offers a path to treat more extended systems without loss of accuracy in the solution of the scattering equations.   The fundamental idea combines properties of the numerical Schwinger variational method, which has proved difficult to apply in multichannel calculations on polyatomic molecules, and the more straightforward interface with electronic structure theory for multichannel treatments of polyatomics offered by the Complex Kohn approach.

Although the literature on the theory of electron-molecule collision processes and molecular photoionization is now large and well established, the results from new experimental methods, in particular  attosecond molecular photoionization~\cite{attosec_nature_report_2014,calegari2014ultrafast} and momentum imaging measurements of dissociative electron attachment (DEA) processes~\cite{Slaughter_DEA_review_2016,Krishnakumar_DEA_CF4_2014,rescigno2016_NH3_dynamics,haxton2011observation}, are challenging the current capabilities of all existing \emph{ab initio} methods. An example is the problem of the physics of dissociative electron attachment to DNA bases~\cite{Kawarai_uracil_2014,McKoy_2008,tennyson_uracil_2009},  in which the molecules are of medium size and for which the more correlated treatments currently possible for small molecules are out of the practical reach of present methods.   It is becoming increasingly urgent for theory to address the highly correlated electronic continuum processes revealed in these classes of experiments when they are applied to moderate-sized polyatomic molecules 

Since it was first proposed, the Kohn variational principle~\cite{kohn_1948} for scattering amplitudes has been applied using trial functions that employ a basis expansion of the scattering wave function in the interaction region combined explictly with free-particle continuum wave functions, so that the trial function will satisfy proper scattering boundary conditions.   Here we replace this form of the trial wave function with a grid-based representation making use of a version of the overset grid approach ~\cite{suhs2002pegasus,meakin1998composite,petersson1999holecutting,chan2002best} that is familiar in the fluid dynamics literature.
For molecules, our overset grids consist of a sparse central grid located
at the center of charge or an atom of interest, together with a set of dense 
subgrids centered on each atom.
An example of a molecular overset grid is shown in
Fig.~\ref{fig:CF4OversetGrid}.  Wave functions are switched between grids using Becke switching
functions~\cite{becke1988multicenter}.
On an overset grid, the wave function is effectively simultaneously expanded in partial waves
about every nuclear center as well as about the center of the coordinate system, and this expansion converges much faster than the familiar 
 single-center expansions that have been used, for example in Schwinger variational calculations~\cite{lucchese1981iterative,lucchese1982studies}.  
 
The central difficulty in applying grid methods to the Kohn variational principle is that the Kohn trial function must explictly satisfy scattering boundary conditions, and in the Complex Kohn approach those must be the complex outgoing wave boundary conditions~\cite{mccurdy_interrelation_1987,miller_jansenopdehaar_1987}.  To construct the trial function explicitly satisfying scattering boundary conditions using only a grid representation, we expand it in functions that are powers of $\hat{G}^+_0\hat{V}$ operating on 
the incoming free-electron scattering function, where $\hat{G}^+_0$ and
$\hat{V}$ represent the free-particle outgoing-wave Green's function and the molecular
potential, respectively.  We find that this choice of basis is equivalent to constructing a Pad\'e approximant to the exact $T$-matrix being approximated by the Kohn variational expression.  The expansion is accomplished numerically as the basis of an iterative Arnoldi solution of the Kohn scattering equations, called ``Born-Arnoldi'' iterations here, that exhibits remarkably fast convergence,  due evidently to the underlying equivalence to the construction of a Pad\'e approximant to the solution.    

 In the following sections we give a complete description of the grid-based complex Kohn method and demonstrate it in calculations on small molecules.  In Sec.~\ref{sec:ComplexKohn} we formulate the Complex Kohn variational method  and describe its  relation to Pad\'e approximants when implemented with the form of the trial function we are employing.  We develop overset grids for molecules in 
Sec.~\ref{sec:OversetGrid}, and we 
describe the Born-Arnoldi procedure in Sec.~\ref{sec:BornArnoldi}. 
In Sec.~\ref{sec:Operations} we describe the implementation of the 
three-dimensional free-particle Green's function and both local and nonlocal potential operations.  These are based on radial subgrids using the finite-element discrete-variable-representation, which
requires the application of subinterval integration
weights to treat the slope discontinuity in the radial Green's functions.
In Sec.~\ref{sec:Results} we present a series of numerical tests of the method. First in Sec.~\ref{sec:NoroTaylor}, with a model problem we show that the method is invariant to using an overset grid that does not match the underlying symmetry of the potential .
Then, in comparisons with electron-molecule scattering calculations using the Schwinger approach we demonstrate the advantages of the overset grids and the
Born-Arnoldi procedure in Secs.~\ref{sec:CH4}~and~\ref{sec:CF4} in static and static-exchange calculations on the
 CH$_4$ and CF$_4$ molecules.

\section{Theory and implementation \label{sec:Theory}}
We begin by briefly describing the Complex Kohn variational method, and the connection to Pad\'e approximants when the trial function is expanded in the basis consisting of $(\hat{G}^+_0 V)^n$ operating on the incoming wave.
We describe the general extension to close-coupling calculations including electron correlation and target response, to highlight 
some of the difficulties with applying the Complex Kohn method to large molecules in 
its previous implementations, and then discuss how they can be overcome by 
using overset grids. After developing the overset grids, we present the Born-Arnoldi
iteration procedure as a quickly converging method to solve the resulting system
of linear equations, containing millions of variables. Finally, we
illustrate the application of the necessary operations on the overset
grids.

\subsection{Complex Kohn variational method and its relation to Pad\'e approximants \label{sec:ComplexKohn}}

The essence of Complex Kohn variational approach can be seen easily in its formulation for a potential scattering problem. The original formulation of the variational principle by Kohn~\cite{kohn_1948} made use of real-valued trial functions and reactance matrix boundary conditions.  Many years later when it was reformulated with complex-valued S-matrix or T-matrix scattering boundary conditions~\cite{miller_jansenopdehaar_1987,mccurdy_interrelation_1987}, a complex symmetric inner product was introduced in place of the usual Hermitian inner product, thereby removing spurious singularities that appeared in numerical implementations of the original reactance matrix form.  To clarify the origin of the complex symmetric representation of the Hamiltonian that has appeared in all implementations of the Complex Kohn variational method to date, we begin with the correct formulation of the Complex Kohn variational principle using the conventional Hermitian inner product.
 
 The Kohn stationary functional, $T^{+\textrm{S}}_{\mathbf{k}',\mathbf{k}}$,  of the trial functions, $\psi_{\mathbf{k}}^{(+)t}$ and $\psi_{\mathbf{k}'}^{(-)t}$,  is
\begin{equation}
T^{+\textrm{S}}_{\mathbf{k}',\mathbf{k}} = T^{+t}_{\mathbf{k}',\mathbf{k}} +(2\pi)^{-\frac{3}{2}} \bracket{\psi_{\mathbf{k}'}^{(-)t}}{\hat{H}-E}{\psi_{\mathbf{k}}^{(+)t}}
\label{eq:kohn3D}
\end{equation}
where the scattering $T$-matrix is labeled by asymptotic momenta, and the scattering states are $\delta_3(\mathbf{k}'-\mathbf{k})$ normalized.  In this paper we will consider the case of electron-molecule scattering where only a single target electronic state is included, i.e. potential scattering from a non-spherically-symmetric non-local potential $\hat{V}$.  In that case, the $T$-matrix used in Eq.~\eqref{eq:kohn3D} has been defined such that
\begin{equation}
T^{+}_{\mathbf{k}',\mathbf{k}} = \bracket{\psi_{\mathbf{k}'}^{0}}{\hat{V}}{\psi_{\mathbf{k}}^{(+)}},
\end{equation}
where $\psi_{\mathbf{k}'}^{0}$ is the unscattered plane wave.
With partial-wave expansions of the scattering states of the form
\begin{equation}
\psi^{(\pm)}_{\mathbf{k}}(\mathbf{r})=\sqrt{\frac{2}{\pi}} \sum_{l,m} i^l   \psi^{(\pm)}_{klm}(\mathbf{r}) Y_{l,m}^* (\hat{\mathbf{k}}),
\label{eq:psipwdef}
\end{equation}
the partial-wave expansion of the $T$-matrix is written as
\begin{equation}
T^{+}_{\mathbf{k}',\mathbf{k}}= \sum_{l',m',l,m} (i)^{l-l'} \frac{2}{\pi} T_{k,l',m',l,m}^+ Y_{l',m'}(\hat{\mathbf{k}}') Y_{l,m}^*(\hat{\mathbf{k}}),
\end{equation}
where
\begin{equation}
T_{k,l',m',l,m}^+  = \bracket{\psi_{kl'm'}^{0}}{\hat{V}}{\psi_{klm}^{(+)}}.
\end{equation}
The partial-wave form of the Complex Kohn variational expression is then
\begin{equation}
T^{+\textrm{S}}_{k,l',m',l,m} = T^{+t}_{k,l',m',l,m} +  \bracket{\psi^{(-)t}_{kl'm'}}{\hat{H}-E}{\psi^{(+)t}_{klm}}.
\label{eq:kohn}
\end{equation}

The traditional Complex Kohn trial function has the form of the sum of contributions of a square integrable expansion basis that describes the collision region and terms that allow it to satisfy scattering boundary conditions,
\begin{equation}
\begin{split}
\psi^{(\pm)t}_{klm}= &\sum_i c_i^{(\pm)} \, \varphi_i(\mathbf{r})+  
\frac{1}{kr}\bigg[  \hat{j}_{l}(kr) Y_{lm}(\hat{\mathbf{r}})   \\
&\qquad \qquad  +\sum_{l',m'}T^{\pm t}_{k,l',m',l,m}\tilde{h}^{\pm}_{l'}(kr)   Y_{l'm'}(\hat{\mathbf{r}})  \bigg]
\end{split}
\label{eq:trialfcn}
\end{equation}
In Eq.~\eqref{eq:trialfcn} the functions $\varphi_i$ are (usually real-valued) square-integrable functions,  $ \hat{j}_{l}(kr)$ denotes the regular Riccati-Bessel function, and $\tilde{h}^{\pm}_l(kr) \xrightarrow{r \rightarrow \infty} \hat{h}^{\pm}_l(kr)$  is a function regular at the origin that becomes the outgoing Riccati-Hankel function asymptotically.  The variational parameters in this linear trial function are the coefficients $c_i^{(\pm)}$ and the $T$-matrix elements $T^{\pm t}_{k,l',m',l,m}$.  Inserting the trial function in Eq.~\eqref{eq:kohn} produces the working expression which can be written compactly in its general form for single channel or multichannel scattering,
\begin{equation}
\mathbf{T}^\textrm{S} =  \mathbf{N}_{00} - \mathbf{N}_{0q}\mathbf{M}^{-1}_{qq}\mathbf{N}_{q0}
\label{eq:kohnworking}
\end{equation}
Here the subscript $0$ denotes the set of regular continuum functions, $\{ \hat{j}_l(kr)  Y_{lm}(\hat{\mathbf{r}}) /kr \}$, and the subscript $q$ denotes the expansion basis   $ \{ \, \{ \varphi_i(\mathbf{r}) \},
\{ \tilde{h}^{\pm}_l(kr)  Y_{lm}(\hat{\mathbf{r}})/kr \} \, \}$.  The first of these matrices is the Born term and has the form 
\begin{equation}
\begin{split}
&(\mathbf{N}_{00})_{l',m',l,m} \\
&\quad \quad= \bracket{ \frac{\hat{j}_{l'}(kr)}{kr}  Y_{l'm'}(\hat{\mathbf{r}})}{\hat{V}}{\frac{ \hat{j}_l(kr)}{kr}  Y_{lm}(\hat{\mathbf{r}})}
\end{split}
\label{eq:matrixels}
\end{equation}
The matrices $\mathbf{N}_{0q}$ and $\mathbf{N}_{q0}$ are similarly defined.  The $\mathbf{M}_{qq}$ matrix elements are brackets of $(\hat{H}-E)$ between functions in the expansion basis.  For all matrix elements in $N$ and $M$,  the $(+)$ functions are used on the right-hand sides of the brackets and the $(-)$ functions are used on the left-hand sides.  This is the Complex Kohn formulation that has no singularities in the matrix inverse, $\mathbf{M}^{-1}_{qq}$, at real scattering energies, and it is equivalent to the ``S-matrix Kohn'' formulation of Miller and Jansen op de Haar~\cite{miller_jansenopdehaar_1987}.   If the basis functions on the left and right are related by the $+\leftrightarrow -$ relationship, i.e complex conjugation and $\vec{k} \leftrightarrow -\vec{k}$, the matrix $\mathbf{M}_{qq}$ is  complex symmetric in these approaches, unlike in the original formulation of Kohn~\cite{kohn_1948}, where the equivalent matrix is real symmetric.

As mentioned in section \ref{sec:intro}, the grid version of the Kohn variational approach must represent the continuum orbitals on the overset grid, but it must also apply the asymptotic boundary conditions in Eq.~\eqref{eq:trialfcn}.  We can do that by expanding $\psi^t$ in a set of functions we construct on the grid by operating with the free-particle Green's function, $\hat{G}^+_0$, which here denotes the Green's function for outgoing boundary conditions $\hat{G}_0^+ \equiv (E-\hat{T} +i\epsilon)^{-1}$, where $\hat{T}$ is the kinetic energy operator.  The grid-based trial function is now,
\begin{eqnarray}
\ket{\psi^{(\pm)t}_{klm}} &=& \ket{\phi^0_{klm}} + \sum_{i=1}^N   c_i \ket{\phi_{i,klm}^{(\pm)}}  \\
\ket{\phi_{i,klm}^{(\pm)}}& \equiv &(\hat{G}_0^{\pm} \hat{V})^i \ket{\phi^0_{klm}}
\label{eq:G0Vbasis}
\end{eqnarray}
with $\phi^0_{klm} =  \hat{j}_{l}(kr)  Y_{lm}/kr $ being the incoming wave.  Now \emph{all} the functions in the expansion of the trial wave function, except for $\phi^0$, satisfy outgoing wave boundary conditions
 ($\phi_{i,klm}^{(+)}$) or incoming wave boundary conditions ($\phi_{i,klm}^{(-)}$) because of the asymptotic form of $\hat{G}^{\pm}_0$.  
 Note that with this choice of basis functions the bra $\bra{\phi_{i,klm}^{(-)}}$ satisfies
 \begin{equation}
 \bra{\phi_{i,klm}^{(-)}} = \bra{\phi^0_{klm}}(\hat{V}\hat{G}_0^{+})^i .
 \end{equation}
 Thus, with this basis the $N$ and $M$ matrices can be written as
 \begin{equation}
 (\mathbf{N}_{l',m',l,m})_{i,j} = \alpha_{i+j}^{(l',m',l,m)}
 \end{equation}
 where
 \begin{equation}
\alpha_{i}^{(l',m',l,m)} \equiv \bracket{\phi^0_{kl'm'}}{\hat{V}\left(\hat{G}^+_0 \hat{V}\right)^i}{ \phi^0_{klm}},
\end{equation}
and 
\begin{eqnarray}
(\mathbf{M}_{l',m',l,m})_{i,j} &=& \bracket{\phi_{i,kl'm'}^{(-)}}{\hat{H}-E}{ \phi_{j,kl'm'}^{(+)}} \nonumber \\
&=&\alpha_{i+j-1}^{(l',m',l,m)}-\alpha_{i+j}^{(l',m',l,m)}
 \label{eq:Mmatrix}
 \end{eqnarray}
where we have used the identity $(E-\hat{T})\hat{G}^{\pm}_0 =1$. In Sec. \ref{sec:Operations} we will discuss the procedure for operating accurately and efficiently with $\hat{G}_0^{\pm}$, but here we can make an important observation about the properties of this basis. 

Nuttall~\cite{nuttall_pade_1967} and Garibotti ~\cite{garibotti_pade_1972} observed that expressions similar to Eq.~\eqref{eq:kohnworking}  could constitute Pad\'e approximants to the equivalent exact expressions corresponding to the Schwinger variational principle -- providing a particular expansion basis is used for the trial function.  We have found that a similar, but not identical,  result applies here.  The power series expansion of the scattered portion of the exact $T$-matrix in the strength parameter, $\lambda$,  has the form for elastic scattering (for simplicity) and dropping the partial-wave superscripts,
\begin{equation}
\begin{split}
t=& ( \phi_0 | \hat{V} (E-\hat{T} - \lambda \hat{V} + i\epsilon)^{-1}\hat{V} |\phi_0 ) \\
=& \sum_{i=0}^\infty \lambda^i \alpha_{i+1}.
\end{split}
\label{eq:exactT}
\end{equation}
The Kohn variational approximation to Eq.~\eqref{eq:exactT} using the basis in Eq.~\eqref{eq:G0Vbasis} is the second term in Eq.~\eqref{eq:kohnworking}, and for elastic scattering has the form
\begin{equation}
t_P^{(N)} =\vec{\alpha}^T \mathbf{M}^{-1} \vec{\alpha}
\label{eq:pade}
\end{equation}
where $\vec{\alpha}=(\alpha_1, \alpha_2,\cdots,\alpha_N)$.
 
The key point is that Eqs.(\ref{eq:pade}) and (\ref{eq:Mmatrix}) have essentially the same form as Nuttall's Eqs.(28-30) that relate the elements of $M_{ij} = \alpha_{i+j-1} -\lambda \, \alpha_{i+j}$ to each other and to the elements of $\vec{\alpha}$ (differing here slightly in the definition of $\alpha_i$) .  Using the same logic used there~\cite{nuttall_pade_1967}  and by Garibotti~\cite{garibotti_pade_1972} we can verify that Eq.~\eqref{eq:pade} produces an $[(N-1)/N]$ Pad\'e approximant to the scattered portion of the $T$-matrix in Eq.~\eqref{eq:exactT}, i.e., a ratio of polynomials,
\begin{equation}
t_P^{(N)}=P_{N-1}(\lambda)/Q_{N}(\lambda)=\sum_{i=0}^{2N-1} \lambda^i \alpha_{i+1} +\mathcal{O}(\lambda^{2N})
\end{equation}
that reproduces the Born expansion of the scattered term in Eq.~\eqref{eq:exactT} to order $\lambda^{2N-1}$, but that has accelerated convergence properties.   The analogous result in the case of the Schwinger expressions investigated in references \cite{nuttall_pade_1967} and \cite{garibotti_pade_1972} and later used  by Lucchese and McKoy~\cite{lucchese_pade_1983} is an $[N/N]$ Pad\'e approximant. Also note that the variational expression given in Eq.~\eqref{eq:pade} is related to the $M^1_{2,3}(VS',S)$ variational expression discussed in ref.~\cite{lucchese_pade_1983} with the appropriate choice of bases.

The analysis for off-diagonal $T$-matrix elements is similar to that in Garibotti~ \cite{garibotti_pade_1972}.   While we know of no general proofs concerning the rate of convergence of the Pad\'e approximants that we effectively construct here, the connection to Pad\'e approximants gives a strong hint as to the origins of the extraordinarily rapid rate of convergence we observe in the grid-based Kohn method demonstrated numerically in Secs. \ref{sec:BornArnoldi} and \ref{sec:CF4} below.

The working equations of the many-electron coupled-channels version of this theory has the form~\cite{mccurdy1989collisions} of  Eq.~\eqref{eq:kohnworking}, but the matrices are defined in terms of the components of the many-electron trial function corresponding to incoming waves in channel $\Gamma$,
\begin{equation}
\begin{split}
\psi^t_{\Gamma'} =& \sum_{\Gamma}  \mathcal{A} \chi_{\Gamma} (\mathbf{r_1} \cdots \mathbf{r_N})\,F_\Gamma(\mathbf{r_{N+1}}) \\
& \qquad +\sum_\mu d_\mu^{\Gamma'} \Theta_\mu(\mathbf{r_1} \cdots \mathbf{r_{N+1}})
\end{split}
\label{eq:manyekohn}
\end{equation}
In previous implementations of the theory, the continuum ``orbitals'' $F_\Gamma(\mathbf{r_{N+1}})$ of this trial function have the form of Eq.~\eqref{eq:trialfcn} but with the $T$-matrix now labeled additionally by channels, $T^{\Gamma,\Gamma'}_{lm,l'm'}$ and incoming waves in one channel only.  The correlated target state functions, $\chi_\Gamma$, and the square-integrable $N+1$ electron correlation terms, $\Theta_\mu$ , form the interface with electronic structure upon which we will comment further below.  The much larger matrix inverse portion of the working expression, $\mathbf{M}^{-1}_{qq}\mathbf{N}_{q0} $ is of course found by solving linear equations, as we will do in Sec. \ref{sec:BornArnoldi} in the grid implementation.

In the previous implementations of the complex Kohn variational method, a
basis of atom-centered Gaussians and central Bessel functions is used to
expand the trial function in Eq.~\eqref{eq:manyekohn}~\cite{rescigno1995incorporation,rescigno1995complex}. 
This approach has been applied successfully to a number of 
multichannel and single-channel electron-molecule scattering problems~\cite{mccurdy1989collisions,lengsfield_methane_1991,parker_H2_close_coup_1991,lengsfield1991electronmolecule,orel1991dissociative,douget_DEA_CH4_2015} 
and photoionization problems~\cite{rescigno1993interchannel,trevisan_imaging_2012,menssen2016molecular,mccurdy2017unambiguous}.
However, it has some drawbacks that limit its applications to larger
systems. Chiefly, the exchange interaction between continuum and bound
electrons is approximated. As more scattering channels are added, or more
correlated target states considered and therefore more of the basis
required to describe bound states, this approximation (called  the ``separable exchange approximation'' since its introduction~\cite{mccurdy1989collisions,lengsfield_methane_1991}) becomes a greater
hindrance to the accurate solution of the complex Kohn equations.
Additionally, at high energies, the rapid oscillation of the
wave functions become more difficult to describe in this basis.

These issues have been addressed in other variational scattering methods
such as the Schwinger variational method by using a single-center grid
expansion of the wave function~\cite{lucchese1981iterative,lucchese1982studies}. 
The Schwinger variational method has the
disadvantage that describing multichannel scattering is much more
difficult than in the complex Kohn variational method, which only
requires a multichannel trial wave function. Furthermore, the use of
single-center grid expansions limits the size of the molecules that can
be considered, as we will show in Section~\ref{sec:Results}. This limitation arises 
because for large molecules with multiple heavy atoms, the cusps in the
wave function at each nuclear position becomes more difficult to resolve
as they move further from the expansion center.  It may require hundreds or thousands of angular grid
points to resolve a nuclear cusp, leading to grids with tens of millions of points. This is the
case even for small molecules such as CF$_4$, as we will show in
Section~\ref{sec:Results}. The solution to these problems lies in the
use of an overset grid.

\subsection{Overset grid \label{sec:OversetGrid}}
The overset grid, pictured for CF$_4$ in Fig.~\ref{fig:CF4OversetGrid},
consists of a central  grid (Carbon-centered in
Fig.~\ref{fig:CF4OversetGrid}) and several smaller but denser subgrids
(Fluorine-centered in Fig.~\ref{fig:CF4OversetGrid}).  The grids we use are combinations of finite-element-method discrete-variable-representation (FEM-DVR) grids for the radial variable in each subgrid~\cite{lill1982discrete,rescigno2000numerical,tao2009gridbased} with Gauss-Chebyshev and Gauss-Legendre quadratures in the angular variables ~\cite{lucchese_1990,gianturco1994calculation}.  
A wave function described everywhere in space can be switched onto the
each grid $g$ using Becke switching functions,
$W_g$~\cite{becke1988multicenter}, commonly used in numerical density functional calculations, that smoothly switch between unity inside the grid and zero outside it, 
\begin{equation}
\Psi(\mathbf{r}) = \sum_g W_g(\mathbf{r}) \Psi(\mathbf{r}) .
\label{eq:SwitchPsi}
\end{equation}
The left-hand and right-hand sides of Eq.~\eqref{eq:SwitchPsi} are equal
because the switching functions sum to unity everywhere in space,
\begin{equation}
\sum_g W_g(\mathbf{r}) = 1 .
\label{eq:SumSwitchUnity}
\end{equation}
The right-hand side of Eq.~\eqref{eq:SwitchPsi} is a sum of
wave functions localized on each grid,
\begin{equation}
\psi_g(\mathbf{r}-\mathbf{r}_{g,0}) = W_g(\mathbf{r}) \Psi(\mathbf{r}) ,
\label{eq:PsiLocal}
\end{equation}
where $\mathbf{r}_{g,0}$ is the origin of grid $g$. The localized functions can be
expanded in local partial waves,
\begin{equation}
\psi_g(\mathbf{r}_g) = \sum_{l,m} \frac{1}{r_g}\psi_{g,lm}(r_g) Y_{lm}(\hat{\mathbf{r}}_g),
\label{eq:PsiLocalPW}
\end{equation}
with $\mathbf{r}_g=\mathbf{r}-\mathbf{r}_{g,0}$ and the $Y_{lm}(\hat{\mathbf{r}}_g)$ are spherical harmonics in the angular coordinates
of each grid.  Furthermore, there is an underlying polynomial basis of normalized Lobato shape functions $\chi_q(r_g)$ connected to the FEM-DVR radial grids \cite{rescigno2000numerical} so that in a direct product basis
\begin{equation}
X_{g,\lambda}(\mathbf{r}_g) = \chi_q(r_g) Y_{lm}(\hat{\mathbf{r}}_g),
\end{equation}
where $\lambda=(q,l,m)$, the localized wave functions can written as a linear combination of this basis as
\begin{equation}
\psi_g(\mathbf{r}_g) = \frac{1}{r_g} \mathbf{X}_{g}(\mathbf{r}_g) \mathbf{c}_g,
\label{eq:PsiBasisPW}
\end{equation}
where the elements of the column vector $\mathbf{c}_g$ are the basis expansion coefficients of the local
function and $\mathbf{X}$ is a row vector of the Lobato shape functions
\begin{equation}
\mathbf{X}_g(\mathbf{r}_g) = \left[X_{g,1}(\mathbf{r}_g),X_{g,2}(\mathbf{r}_g),...,X_{g,N}(\mathbf{r}_g)\right]
\label{eq:XVec}
\end{equation}
Our choice of the angular quadratures allow an efficient transformation between the grid and spherical harmonic representations~\cite{gianturco1994calculation}.

We will discuss in detail the operations necessary for the Born-Arnoldi
scheme in Sec.~\ref{sec:Operations} but here we consider how operators are represented in the overset grid approach.  A three-dimension function, 
$\Psi(\mathbf{r})$, is defined by its value on the grid points $\mathbf{r}_g$ on all grids labeled by the different values of $g$.  For a local operator, $\hat{O}=\hat{O}(\mathbf{r})$, we have
\begin{equation}
(\hat{O}\Psi)(\mathbf{r}) = \hat{O}(\mathbf{r})\Psi(\mathbf{r}).
\end{equation}
For non-local operators, e.g. exchange operators and Green's functions, $\hat{O}$ operating on a set of local function $\hat{O}\psi_g$ can be
calculated using the basis representation of the operator $\mathbf{O}_g$ so that
\begin{equation}
(\hat{O}\psi_g)(\mathbf{r}_{g}) = \frac{1}{r_g}\mathbf{X}(\mathbf{r}_{g})\mathbf{O}_g\mathbf{c}_g,
\label{eq:OonGrid}
\end{equation}
The off-grid contributions must be calculated by interpolation or extrapolation.
The FEM-DVR and spherical harmonic expansions lend themselves to interpolating accurately due to the underlying
orthogonal polynomial bases.
If the maximum radius of the grid in region $g$ is $r_{g,x}$, then the off-grid points are obtained from a matrix
transform~\cite{boyd2001chebyshev}, 
\begin{equation}
\begin{split}
&(\hat{O}\psi_g)(\mathbf{r}-\mathbf{r}_{g,0}) \\
&\quad= \begin{cases}
\frac{1}{|\mathbf{r}-\mathbf{r}_{g,0}|}\mathbf{X}_g(\mathbf{r}-\mathbf{r}_{g,0})\mathbf{O}_g\mathbf{c}_g & | \mathbf{r}-\mathbf{r}_{g,0}| < r_{g,x} \\
\mathbf{Z}_g(\mathbf{r}-\mathbf{r}_{g,0})\mathbf{O}_g^{a}\mathbf{c}_g& \textrm{ otherwise}
\end{cases} ,
\label{eq:OoffGrid}
\end{split}
\end{equation}
where $\mathbf{Z}_g$ is a row vector of the asymptotic forms of the operator $O$ outside the grid $g$.  For example, in the case
of the $1/r_{12}$ operator, the asymptotic forms for a system with no symmetry would be
\begin{equation}
\begin{split}
\mathbf{Z}_g(\mathbf{r}_g)& = \left[\frac{1}{r_g}Y_{00}(\hat{\mathbf{r}}_g),\frac{1}{r_g^2}Y_{11}(\hat{\mathbf{r}}_g),\frac{1}{r_g^2}Y_{10}(\hat{\mathbf{r}}_g),\right.\\
&\left.\frac{1}{r_g^2}Y_{1,-1}(\hat{\mathbf{r}}_g),\frac{1}{r_g^3}Y_{2,-2}(\hat{\mathbf{r}}_g),\frac{1}{r_g^3}Y_{2,-1}(\hat{\mathbf{r}}_g),\ldots\right].
\end{split}
\label{eq:ZVecExample}
\end{equation}
The operation on the total wave function can then be determined by
\begin{equation}
(\hat{O}\Psi)(\mathbf{r}) = \sum_{g} (\hat{O}\psi_{g})(\mathbf{r}-\mathbf{r}_{g,0}).
\label{eq:OTotal}
\end{equation}

In the single channel calculations that we present here the matrix inverse portion of 
of the Complex Kohn working expression in Eqs.(\ref{eq:kohnworking})  and (\ref{eq:matrixels}) is equivalent to the solution of  a driven equation
\begin{equation}
(E-\hat{H}) \ket{\Psi^{sc}} = \hat{V} \ket{\frac{\hat{j}_{l}(kr)}{kr}Y_{lm}(\hat{\mathbf{r}})}
\label{eq:ComplexKohnDriven}
\end{equation}
of the Schrodinger equation, which when expressed directly on the overset grid becomes a set of linear equations for the quantity $\Psi^{sc} = \mathbf{M}^{-1}_{qq}\mathbf{N}_{q0} $.   The coordinate $\mathbf{r}$ in
Eq.~\eqref{eq:ComplexKohnDriven} is defined in the central grid of the 
overset grid.
Three problems arise in solving the the complex Kohn driven
equations, Eq.~\eqref{eq:ComplexKohnDriven}, on the grid:  i)  outgoing wave scattering boundary conditions must be imposed on 
the solution,  ii) the sets of linear equations on the overset grids can be on the order
of millions of variables, and iii) interpolating the derivatives
required by the kinetic energy operator in $\hat{H}$ between grids as
required in Eq.~\eqref{eq:OTotal} is numerically problematic due to the
oscillations of the DVR representation of the derivative operator between grid points.
The use of the Born-Arnoldi iterative basis defined in Eq.~\eqref{eq:G0Vbasis} and the Kohn variational expression given in Eq.~\eqref{eq:kohnworking} will be seen to overcome these problems.

\subsection{Born-Arnoldi iteration \label{sec:BornArnoldi}}
By building a basis to solve Eq.~\eqref{eq:ComplexKohnDriven} using the Born
series, $(\hat{G}^+_0\hat{V})^n\hat{j}_lY_{lm}/kr$, we have a physically motivated basis 
that automatically satisfies the outgoing wave scattering
and that is quickly convergent, evidently because of its equivalence to a Pad\'e approximant. 
Additionally, as we will show in this section, this
basis eliminates the need for interpolating the action of the kinetic 
energy operator $\hat{T}\Psi$, significantly reducing the numerical error
in the overset grid representation.

\begin{figure}
\includegraphics[width=0.95\linewidth]{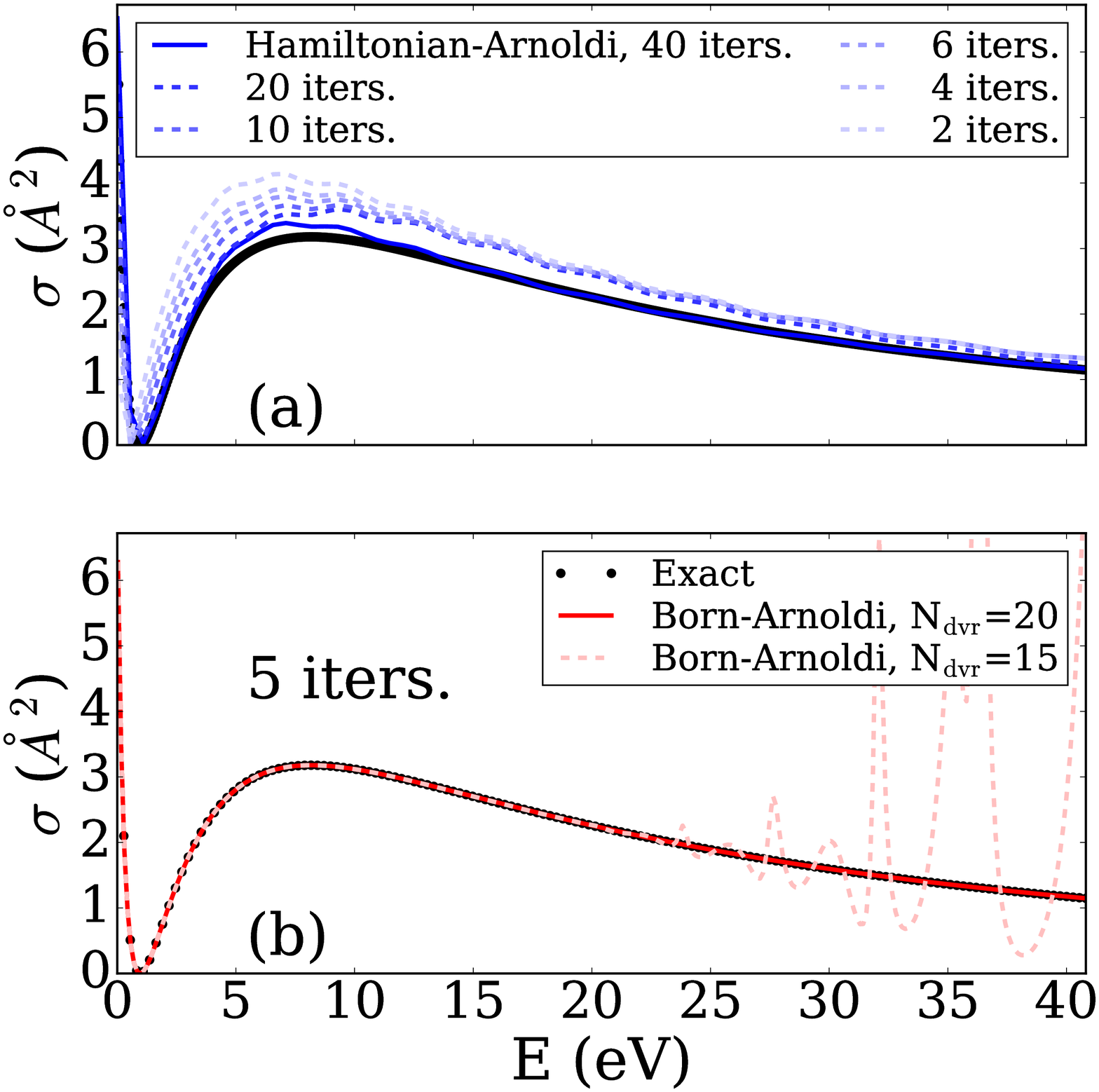}
\caption{(Color online)
Partial scattering cross sections for s-waves for the spherical potential
$V=-3e^{-r}$.
The Hamiltonian-Arnoldi iterates of $E-\hat{H}$ (upper panel) converge slowly to the 
exact result.
The Born-Arnoldi iterates (lower panel) converge at only 5
iterations, including in the case where the order of the DVR in each finite element is increased to represent the 
scattered wave at higher energies. 
\label{fig:ArnoldiConvergence}}
\end{figure}

We aim to solve the driven equation of the complex Kohn variational
method,  Eq.~\eqref{eq:ComplexKohnDriven} with $E=k^2/2$, in order to 
construct the $T$-matrix elements $T_{lm,l'm'}$.
To compare to a more standard Arnoldi iterative approach, we solved those equations with a basis 
of Arnoldi iterates of $E-\hat{H}$ operating repeatedly on the right 
hand side of Eq.~\eqref{eq:ComplexKohnDriven} and applied a cutoff to prevent the results from
reaching the end of the grid, and then added a 
single function on the grid that satisfies outgoing wave boundary conditions 
in each partial wave.  This procedure is analogous to the construction of the 
the traditional Kohn trial function in Eq.~\eqref{eq:trialfcn}.   In Fig.~\ref{fig:ArnoldiConvergence} we 
demonstrate that this procedure, which is labeled ``Hamiltonian-Arnoldi'',  converges quite slowly.  
We have also found that this procedure converges even more slowly at higher energies.

The issues of slow Arnoldi convergence and application of the proper
boundary conditions are both be addressed by using iterates of the
Born series, $(\hat{G}_0^+\hat{V})^n\hat{j}_{l}Y_{lm}/kr$, beginning with $n = 1$.
As discussed Sec.~\ref{sec:ComplexKohn}, using this basis is equivalent to constructing
an $[(N-1)/N]$ Pad\'e approximate to the elements of the scattered portion of the $T$-matrix. 
In addition, the application of the  outgoing free-particle Green's
function $\hat{G}_0^+$ ensures that every member of the basis satisfies outgoing wave boundary
conditions.
The zeroth iterate is set to $\phi_0=\hat{j}_{l}Y_{lm}/kr$ ($\phi_0$ does not
have the proper boundary conditions and is not used to solve
Eq.~\eqref{eq:ComplexKohnDriven}).  Subsequent orthogonal
iterates are generated as follows,
\begin{eqnarray}
\tilde{\phi}^+_{k+1} &= &\hat{G}_0^+\hat{V}\phi^+_k \label{eq:BornArnoldiG0V} ,\\
\bar{\phi}^+_{k+1} &= &\tilde{\phi}^+_{k+1} - (\tilde{\phi}^+_{k+1},\phi^+_j)\phi^+_j \qquad \forall j \leq k \label{eq:BornArnoldiOrthogonalize} ,\\
\phi^+_{k+1} &= &\bar{\phi}^+_{k+1}/(\bar{\phi}^+_{k+1},\bar{\phi}^+_{k+1})^\frac{1}{2} \label{eq:BornArnoldiNormalize} .
\end{eqnarray}
The products $(\phi^+_1,\phi^+_2)$ in
Eqs~\eqref{eq:BornArnoldiG0V}~--~\eqref{eq:BornArnoldiNormalize} are
symmetric products rather than Hermitian inner products,
\begin{equation}
(\phi^+_1,\phi^+_2) = \int \phi^+_1 \phi^+_2 d^3r .
\label{eq:SymmetricProduct}
\end{equation}
Note that this form of the inner product for functions expanded in real-valued symmetry-adapted harmonics, as is done in the actual calculations, is equivalent to the Hermitian inner product with the $(-)$ form of the functions on the left side and the $(+)$ form on the right, i.e. $\braket{\phi^-_1}{\phi^+_2}$.

The use of the Born iterates also eliminates the necessity in the
application of $E-\hat{H}$ to operate numerically with the kinetic energy operator.
The matrix elements of $E-\hat{H}$ in this basis are easily constructed using 
the fact that $G^+_0$ is the Green's function of $E-\hat{T}$.  All the terms in this basis, 
contributing to matrix elements $\bracket{\phi^-_j }{ E-\hat{H}}{ \phi^+_k}$ in the representation of 
Eq.~\eqref{eq:ComplexKohnDriven},
involve the operation
\begin{align}
\nonumber
(E-\hat{H})\hat{G}^+_0\hat{V}\phi^+ &= (E-\hat{T}-\hat{V})\hat{G}^+_0\hat{V}\phi^+ \\
\nonumber
&= (\hat{I}-\hat{V}\hat{G}^+_0)\hat{V}\phi^+ \\
&= \hat{V}(\phi^+ - \hat{G}^+_0\hat{V}\phi^+) . \label{eq:BornArnoldiEminusH}
\end{align}
Since the quantities $\phi^+$ and $\hat{G}^+_0\hat{V}\phi^+$ are calculated for all $\phi^+$ functions 
in the basis, they can be simply combined to construct the result of $E-\hat{H}$ operating on any $\phi^+_k$.  
The calculation of  the matrix element $\bracket{\phi^-_j}{E-\hat{H}}{ \phi^+_k}$ 
is completed by using the quadratures and switching functions for each grid.

\subsection{Operating with $\hat{G}^+_0$ and $\hat{V}$ \label{sec:Operations}}
To complete the algorithm for solving the Complex Kohn equations we require efficient ways with
which to operate with $\hat{G}^+_0$ and the potential on the overset grid.  The potential energy
operator in each channel and the coupling potentials between channels are in general combinations of direct opreators and nonlocal exchange operators.   The essential observation is that since both the free-particle Green's function, $\hat{G}^+_0$ and the potential energy of electron repulsion $1/|\mathbf{r}-\mathbf{r'}|$ are translationally invariant, we can transform to the single-center expansion around the center of each subgrid when operating on that grid.  Exploiting that translational invariance efficiently requires the fast transformation between quadrature points and the spherical harmonic expansion developed earlier~\cite{gianturco1994calculation}, and this is the central idea that allows the overset grid method to effectively expand the wave function around all the nuclear centers and the center of the master grid simultaneously.

We use the integral form of the free-electron Green's function
$\hat{G}^+_0$, operating in the partial wave basis of
Eq.~\eqref{eq:PsiLocalPW},
\begin{align}
\nonumber
\hat{G}^+_0\psi_g(\mathbf{r}_g) = & \sum_{l,m} -\frac{2}{kr_g}Y_{lm}(\hat{\mathbf{r}}_g)\times\\
& \int \hat{j}_l(kr_{<,g})\hat{h}^+_l(kr_{>,g})\psi_{g,lm}(r_g') dr_g' .
\label{eq:G0Integral}
\end{align}
In Eq.~\eqref{eq:G0Integral}, $\hat{h}^+_l$ is the outgoing Riccati-Bessel
function, the arguments of $\hat{j}_l$ and $\hat{h}^+_l$ depend on whether $r_g < r_g'$,
and $\hat{G}^+_0$ is diagonal in the partial-wave basis.
The local wave functions on each grid are expanded in terms of the FEM-DVR local radial basis
functions $\chi_q(r)$ and the wave function evaluated at Gauss-Lobatto
quadrature points $r_q$,
\begin{equation}
\psi_{g,lm}(r_g) = \sum_q \psi_{g,lm}(r_q) \chi_q(r_g) .
\label{eq:FEMDVRExpansion}
\end{equation}
At the FEM-DVR element boundaries $R_b$, the integrals in
Eq.~\eqref{eq:G0Integral} can be performed using the Gauss-Lobatto
quadrature with weights $w_q$,
\begin{equation}
(\hat{G}^+_0\psi_{g,lm})(R_b) = -\frac{2}{k}\sum_q w_q\hat{j}_l(kr_<)\hat{h}^+_l(kr_>)\psi_{g,lm}(r_q) ,
\label{eq:G0QuadratureBoundary}
\end{equation}
where $r_<$ and $r_>$ are the minimum and maximum, respectively, of
$R_b$ and $r_q$.

More care must be taken~\cite{basden1988high}  when evaluating the integral in
Eq.~\eqref{eq:G0Integral}, $(\hat{G}^+_0\psi_{g,lm})(r_i)$,  at quadrature points  $r_i$ within an FEM-DVR
element because of the slope discontinuity at $r = r'$.
Naively, the sum in Eq.~\eqref{eq:G0QuadratureBoundary} might be cut off
at $q=r_i$.
This approximation corresponds, however, to integrating a function across the entire
boundary region that is represented by a polynomial whose values at the
quadrature points drop suddenly from the integrand values of
Eq.~\eqref{eq:G0Integral} before $r_i$ to zero following $r_i$.
The resulting polynomial approximation is highly oscillatory and only on very dense FEM-DVR
grids does it represent the integrand accurately.

A far better choice for evaluating Eq.~\eqref{eq:G0Integral} at the interior
points is obtained by integrating the basis functions, $\chi_q(r)$,  of
Eq.~\eqref{eq:FEMDVRExpansion}, which are the cardinal functions of the 
DVR representation,  to obtain integration weights, $\Omega_{q',q}$, adapted to integrating over the subintervals ending
at the quadrature point $r_{q'}$.  For the example of the  finite element beginning at $0$ and ending at $R_b$ we define the adapted weights, 
\begin{align}
\Omega_{q',q}^< &= \int_0^{r_{q'}} \chi_q(r) dr \, , \label{eq:PartialIntegrationWeight} \\
\Omega_{q',q}^> &= \int_{r_{q'}}^{R_b} \chi_q(r) dr \, . \label{eq:PartialIntegrationWeightgreater} 
\end{align}
These weights can be calculated exactly using the same order DVR that defines the cardinal functions
$\chi_q(r)$ by rescaling the integration variable in, e.g, 
Eq.~\eqref{eq:PartialIntegrationWeight},
\begin{align}
\Omega_{q',q}^< = \int_0^{R_b} \chi_q \left(\frac{r_{q'}}{R_b}r\right) \frac{r_{q'}}{R_b} dr . \label{eq:PIWRescaled}
\end{align}
and evaluating the basis functions $\chi_q$ at the rescaled points.
The Green's function result at the interior points $r_i$ are therefore
given as a matrix-vector product of the adapted weights and the
wave functions evaluated across the entire element,
\begin{equation}
\begin{split}
(\hat{G}^+_0\psi_{g,lm})(r_i)= &\,\hat{h}^+_l(kr_i)\sum_q \Omega_{i,q}^< \hat{j}_l(k r_q) \psi_{g,lm}(r_q) \\\
& \quad  +  \hat{j}_l(kr_i)\sum_q \Omega_{i,q}^> \hat{h}^+_l(k r_q) \psi_{g,lm}(r_q) 
\end{split}
\label{eq:G0InteriorPoint}
\end{equation}
producing a significantly more accurate result than truncating the quadrature at
the interior point.

After calculating the on-grid result of the Green's function, we must
interpolate it to obtain the off-grid result,
\begin{equation}
\begin{split}
&(\hat{G}^+_0\psi_g)(\mathbf{r}_{g'}) = -\frac{2}{kr_g[\mathbf{r}_{g'}]} \\
&\qquad \times \sum_{q,lm}(\hat{G}^+_0\psi_{g,lm})(r_q) \chi_q(r_g[\mathbf{r}_{g'}]) Y_{lm}(\hat{\mathbf{r}}_g[\mathbf{r}_{g'}]) .
\end{split}
\label{eq:G0OffGrid}
\end{equation}
The interpolants here are again calculated using the cardinal functions for the radial variables and spherical harmonics for the angular variables.

We must also calculate the operation of the potential $\hat{V}$.
For local potentials, e.g., model potentials for which results are shown in 
Figs.~\ref{fig:ArnoldiConvergence} and \ref{fig:NoroTaylor}, or the Coulomb potential,  we calculate the potential directly
on the coordinates of each grid,
\begin{equation}
\left(\hat{V}\psi_g\right) (\mathbf{r}_g) = V(\mathbf{r}_g)  \psi_g(\mathbf{r}_g) .
\label{eq:ExponentialPotential}
\end{equation}
The nuclear potentials for each atom are such a local potentials, but they are singular at the origin of each subgrid, behaving as $-Z/r$.
The radial functions $\psi_{g,lm}(r)$ in Eqs.~(\ref{eq:FEMDVRExpansion}) and~(\ref{eq:G0QuadratureBoundary})
are $r$ times the complete radial function and tend to 
zero at the grid origins.  We therefore use
L'H\^opital's rule to determine $\hat{V}\psi_g$ at the grid origins,
\begin{align}
(\hat{V}_{\textrm{nuc}}\psi_{g,lm})(r_g\ne0) &= -\frac{Z}{r_g} \psi_{g,lm}(r_g) \, , \label{eq:NuclearPotential} \\
(\hat{V}_{\textrm{nuc}}\psi_{g,lm})(0) &= -Z\left.\frac{\partial \psi_{g,lm}}{\partial r_{g}}\right\vert_{r_g=0} \, . \label{eq:NuclearPotentialOrigin}
\end{align}
Eq.~\eqref{eq:NuclearPotentialOrigin} is only evaluated for the nucleus
at the center of the grid $g$, and the derivative $\chi_q'(0)$ is evaluated using the analytic 
derivative of the FEM-DVR basis functions which include the cardinal function that is nonzero at the origin.

The static Coulomb potential due to an electron in an occupied orbital
$\varphi$ is a non-singular local potential,
\begin{equation}
\left(\hat{J}\{\varphi\}\psi_g\right) (\mathbf{r}_g) = \int\frac{\vert\varphi(\mathbf{r}')\vert^2}{\vert\mathbf{r}'-(\mathbf{r}_g+\mathbf{r}_{g,0})\vert} \psi_g(\mathbf{r}_g)\,  d^3r' .
\label{eq:StaticPotential}
\end{equation}
The molecular orbitals from, for instance, a Hartree-Fock calculation in
a Gaussian basis can be used to provide a static Coulomb potential.
In this case, the integrals of Eq.~\eqref{eq:StaticPotential} evaluated
at each grid point on the overset grid are Gaussian nuclear
electrostatic potential integrals.
These can be computed by standard Gaussian integral libraries in widely
available quantum chemistry software, and we use the \texttt{LIBINT}
integral library~\cite{Libint2} to compute them.

The exchange potential due to an electron in an orbital $\varphi$ is an
example of a non-local potential, and we calculate the exchange
potential using a mechanism similar to the Green's function operation.
The $1/r_{12}$ integrals of the exchange operator can be calculated in
the partial wave basis using the spherical multipole expansion of the
Coulomb interaction,
\begin{equation}
\begin{split}
&\left(\hat{K}\{\varphi\}\psi_g\right) (\mathbf{r}_g) \\
&\quad= \varphi(\mathbf{r}_g)\sum_{l,m}\frac{4\pi}{2l+1}Y_{lm}(\hat{\mathbf{r}}_g)\int\frac{r_{<,g}^l}{r_{>,g}^{l+1}}(\varphi\psi_g)_{lm}(r_g')dr_g' ,
\end{split}
\label{eq:ExchangePotential}
\end{equation}
where $(\varphi\psi_g)_{lm}$ has an overall multiplicative factor of $r_g^2$ coming from the product of two partial-wave radial functions.
Eq.~\eqref{eq:ExchangePotential} is analogous to
Eq.~\eqref{eq:G0Integral} with ${4\pi}/{(2l+1)}$ replacing
$-{2}/{k}$, $r_{<,g}^l$ replacing $\hat{j}_l(kr_{<,g})$, and
$r_{>,g}^{-(l+1)}$ replacing $\hat{h}^+_l(kr_{>,g})$.
The action of any exchange interaction operator is evaluated in the same manner as the Green's function, with these
constants and functions substituted.  

We thus exploit the translational invariance of the underlying operator, $1/r_{12}$, 
to allow the use of a separate single center expansion about the origin of each subgrid.  The combination of these local
expansions and the adapted weights for integrations like that in Eq.~(\ref{eq:G0QuadratureBoundary}) produces an accurate representation on the overset grid of the action of both $\hat{G}^+_0$ and the local and exchange portions of $\hat{V}$.

\section{Numerical results \label{sec:Results}}
We have implemented the complex Kohn variational method on overset 
grids using the computational infrastructure of the electron-molecule scattering code suite
\texttt{ePolyScat}~\cite{gianturco1994calculation,natalense1999crosssection} originally developed for
calculations using the Schwinger approach. This implementation allows interfacing with a number of quantum
chemistry code suites including Molpro~\cite{werner2012molprowires,werner2012molpro} and Gaussian~\cite{g03} to provide 
molecular orbitals from Hartree-Fock calculations and other target electronic structure information.
Below we discuss the application to both model potentials and to single-channel molecular static (Coulomb operators only) and static-exchange approximations.

\subsection{One-dimensional spherical potential \label{sec:NoroTaylor}}

\begin{figure}
\includegraphics[width=0.95\linewidth]{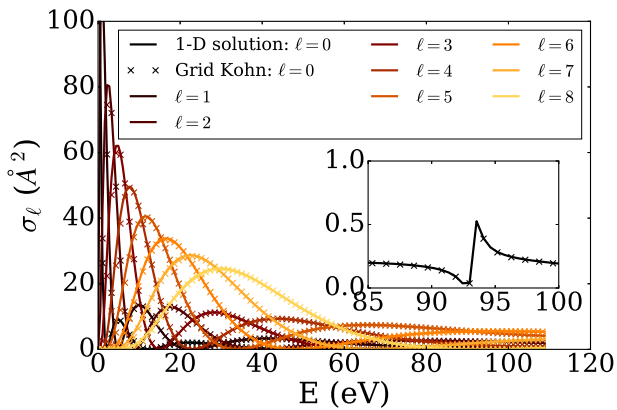}
\caption{(Color online)
The cross sections $\sigma_\ell$ in each angular momentum channel $\ell$
are given for the Noro-Taylor potential $V=\frac{15}{2}r^2e^{-r}$.
The solid lines are numerically exact results calculated using exterior complex
scaling, and the x's are the results calculated using the Complex Kohn
method on the CH$_4$ overset grid.
Even though the potential is peaked at the subgrid locations, the grid-based 
Kohn method is accurate even in the difficult region of the narrow
s-wave resonance (inset).
\label{fig:NoroTaylor}}
\end{figure}

An initial test of the overset grid method is to use a master grid with subgrids that break 
spherical symmetry on 
a spherical potential scattering problem. 
The Noro-Taylor potential~\cite{norotayor_1980}, in atomic units,
\begin{equation}
V(r) = \frac{15}{2}r^2e^{-r} ,
\label{eq:NoroTaylor}
\end{equation}
is 0 at $r=0$, rises to a peak of 110.48 eV at $r=2$ bohr, which is approximately the C-H bond distance in methane, and falls to about 0.93 eV at $r=10$ bohr. In $s$-wave scattering this potential supports a narrow resonance near a scattering energy of 95.2\,eV.
Using the tetrahedral CH$_4$ geometry to build an overset grid, we
observed that the overset grid does not modify the scattering amplitudes of 
the Noro-Taylor potential for any partial wave, nor does it suffer from any 
apparent numerical pathologies associated with the obvious linear dependence of the 
underlying spectral basis in the overlapping subgrids and master grid.
That result is shown in Fig.~\ref{fig:NoroTaylor}.

\subsection{CH$_4$ static Coulomb and static-exchange potentials \label{sec:CH4}}

The central difficulty in the application of  single-center expansions to the solution
of scattering problems is resolving the
nuclear potentials of atoms not at the center of the coordinate system,
which can require values of $\ell$ in the hundreds even for relatively small systems.
This problem of course becomes more severe as heavy
atoms are added far from the expansion center, as we will find in Sec.~\ref{sec:CF4}.  In
the examples we explore here, we will compare the grid-based Complex Kohn approach with 
the numerical Schwinger method which relies on a single center expansion and outward radial integration
beyond the range of the exchange potential, seeking convergence of the integral cross section in all cases to within
0.01\,\AA$^2$.

\begin{figure}
\includegraphics[width=0.95\linewidth]{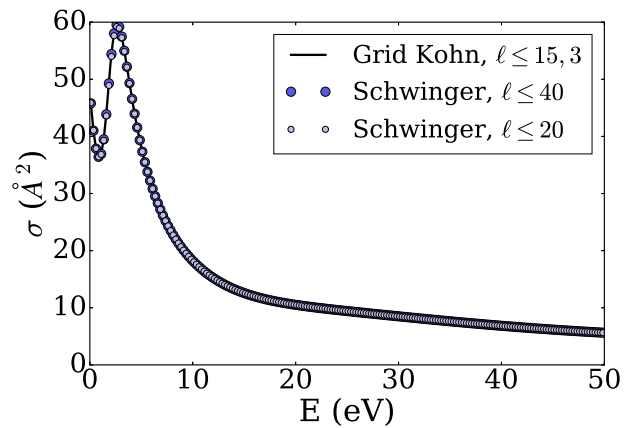}
\caption{(Color online)
Integral cross section for electron-molecule scattering from 
the static potential of CH$_4$ as a function of energy:
Single-center expanded Schwinger variational method  result 
shown with partial waves up to $\ell=40$ (larger dark blue circles) and
$\ell=20$ (smaller light blue circles).
Overset grid complex Kohn method cross sections shown as a solid
black line.
Partial waves only up to $\ell=15$ on the central grid and $\ell=3$ on the
subgrid are required for similar convergence in the complex Kohn method.
\label{fig:CH4Static}}
\end{figure}

We begin by exploring the  cross sections for
scattering from the static Coulomb potential of CH$_4$, calculated using
a STO-3G basis~\cite{hehre1969self,collins1976self}.
We used the Gaussian package~\cite{g03} to calculate the 
Hartree-Fock molecular orbitals.  In Fig.~\ref{fig:CH4Static}, we show the cross sections for the CH$_4$
static potential for both the single-center expanded Schwinger variational
method and the Complex Kohn method on the overset grid.
The single-center expansion is converged using partial waves up to
$\ell=40$, and the grid-based  Kohn method reproduces that result using only
$\ell \le 15$.    

\begin{figure}
\includegraphics[width=0.95\linewidth]{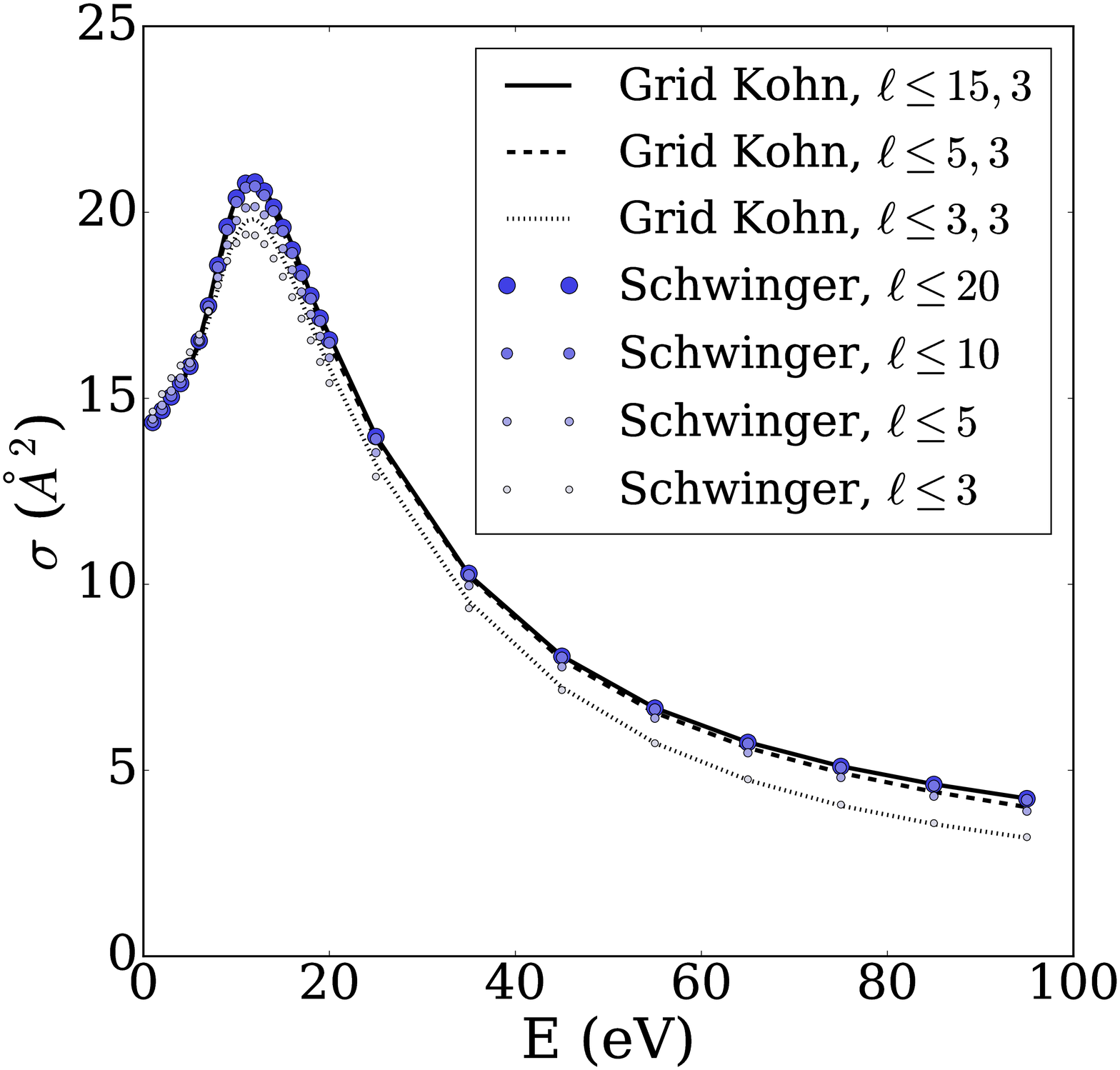}
\caption{(Color online)
The total cross sections $\sigma$ for electron-molecule scattering from 
the static-exchange potential of CH$_4$ are given as a function of energy.
We make a comparison between the single-center Schwinger variational
method (blue circles) from $\ell \leq 3$ (smallest) to $\ell \leq 20$
(largest), and the overset grid complex Kohn method for $\ell \leq 3$
(dotted line), $\ell \leq 5$ (dashed line), and $\ell \leq 15$ (solid
line) on the central grid and $\ell \leq 3$ on the subgrids.
\label{fig:CH4StaticExchange}}
\end{figure}

We show the analogous calculation of the cross sections for the CH$_4$ static-exchange potential
in Fig.~\ref{fig:CH4StaticExchange}. 
The potentials due to CH$_4$ are mostly spherical due to the small
perturbations of the hydrogen atoms, so both the single-center Schwinger
variational method and the overset grid complex Kohn method converge
quickly.
For the Schwinger method, expanding in partial waves with $\ell \leq 20$
and $\ell \leq 10$ give a similar result, while $\ell \leq 5$ and $\ell
\leq 3$ depart considerably from the converged values.
The overset grid complex Kohn method converges more rapidly, converging at
$\ell \leq 5$ until an energy of 50-60\,eV, and completely converging with
$\ell \leq 15$.
While the CH$_4$ calculations test that the method is viable, a more
non-spherical potential is necessary to show that it is a significant
improvement over single-center methods.

\subsection{CF$_4$ static Coulomb and static-exchange potentials \label{sec:CF4}}

CF$_4$ presents a more serious challenge for single-center expansion methods
because of the $Z=9$ F atoms located 1.315\, \AA \,  from the carbon center.
While for CH$_4$, 1,681 partial waves with $\ell \le 40$ were required to
resolve the static Coulomb potential, for CF$_4$ the number of partial
waves required increases dramatically.
For $\ell \le 100$ (10,201 partial waves), the change in many calculated cross 
sections slowed to about 0.1\,\AA$^2$ per increase of $\ell_{\mathrm{max}}$ by 20.
However, as we will show, complete convergence is not
reached even by $\ell_{\mathrm{max}}=200$ (40,401 partial waves).
In contrast, the Complex Kohn method on the overset grid converges to
0.01\,\AA$^2$ between $\ell=35$ and $\ell=45$.
Those values of $\ell$ correspond to 1,296-2,116 partial waves on the central grid and 16 partial
waves on each subgrid.

The total numbers of grid points on the union of the overset grids is therefore in the millions 
for the Complex Kohn calculations despite the more rapid convergence with partial waves in the master
and subgrids.  
Nonetheless, as demonstrated in Table~\ref{table:G0VConvergence}, the number of 
Born-Arnoldi iterations remains less than about 25 even when the number of grid points is doubled.  The 
number of iterations is effectively the number of basis functions in the expansion of the trial wave function in 
Eq.(\ref{eq:G0Vbasis}).  At resonances slightly more iterations can be required, but we observe similar convergence 
properties in the Born-Arnoldi iterations in 
all the calculations reported in  Sec.~\ref{sec:Results}.  The origin of this remarkable numerical behavior is 
evidently the underlying connection to an $[(N-1)/N]$ Pad\'e approximant to the scattered part of the $T$-matrix discussed 
in Sec.~\ref{sec:ComplexKohn}.

\begin{table}
\caption{ \label{table:G0VConvergence}
Convergence of the Born-Arnoldi iterations for electron-CH$_4$ and CF$_4$ scattering 
for different sizes of grids.  $\ell_{max} = \ell_1/\ell_2$ denotes $\ell_1$ maximum
angular momentum on master grid, and $\ell_2$ on subgrids.    All cases refer
to the T$_2$ symmetry component in the static exchange approximation at 6 eV.            
}
\begin{ruledtabular}
\begin{tabular}{ccccc}
             &$\ell_{\mathrm{max}}$ & grid points &     &Born-Arnoldi iterations                 \\
\hline
CH$_4$    &     15/3       &      1,105,920 &     & 9  \\
CH$_4$   &      5/3          &     366,720    &    &  11 \\
CF$_4$ &15/3  &  1,228,800& & 24 \\
CF$_4$ & 35/3  &   2,352,000 & & 22 \\                                                                                                                                          
\end{tabular}
\end{ruledtabular}
\end{table}

\begin{figure}
\includegraphics[width=0.95\linewidth]{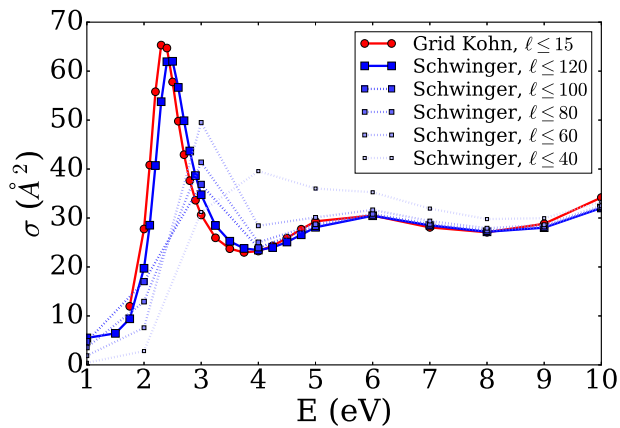}
\caption{(Color online)
A comparison of the cross sections for the static Coulomb potential of
CF$_4$.
The complex Kohn method on the overset grid (red circles, solid line)
is shown with $\ell_{\mathrm{max}}=15$ in the central grid and
$\ell_{\mathrm{max}}=3$ in the subgrids.
Results with the single-center expanded Schwinger variational
method at various $\ell_{\mathrm{max}}$'s are shown (fading purple squares, dotted
lines), including an approximately converged reference line at
$\ell_{\mathrm{max}}=120$ (blue square, solid line).
The single-center results converge slowly with the number of partial
waves, while the overset grid results are largely converged with very few partial waves.
\label{fig:CF4Static}}
\end{figure}

In Fig.~\ref{fig:CF4Static}, we show the cross sections for the CF$_4$
static potential, using the single-center  Schwinger variational
method and the Complex Kohn method on the overset grid.
From $\ell=60$ to $\ell=120$, there is a significant change in the
results obtained using the single-center expansion.
Resolving the static potential is particularly difficult for the
single-center expansion as the slow convergence of the result shows.
The overset grid method, on the other hand, is converged at the small number of
partial waves given by $\ell=25$ in the central grid and $\ell=3$ in the
subgrid.
By expanding the nuclear potentials in the spherical polar coordinates of their
proper centers, we dramatically reduce the computational effort.

\begin{figure}
\includegraphics[width=0.95\linewidth]{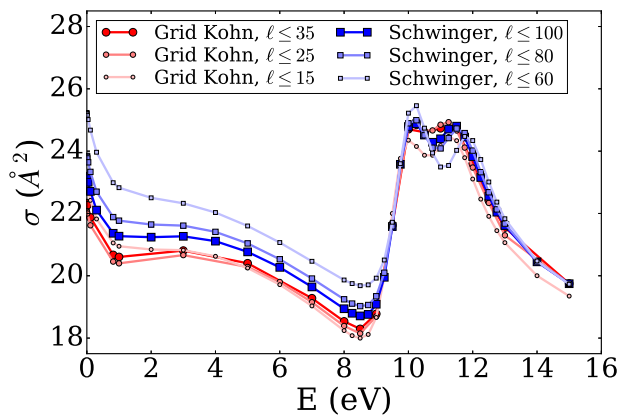}
\caption{(Color online)
Comparison of the cross sections for the static-exchange potential of
CF$_4$. 
The grid Kohn method (red circles, various fading) are given for 
subgrids with $\ell=3$ and central grid with $\ell=$15, 25, and 35.
The non-resonance features match for all three, while the resonance
features converge at $\ell=25$.
The single-center expanded Schwinger method (blue squares,
various fading) is shown for $\ell=$60, 80, and 100 for which some
of the features are not converged even for maximum $\ell=$200.
\label{fig:CF4StaticExchange}}
\end{figure}

In Fig.~\ref{fig:CF4StaticExchange}, we compare the same methods for the
static-exchange potential of CF$_4$ at the HF/6-31G*~\cite{hehre1972self,hariharan1973influence,francl1982self}
level of theory.
The static-exchange potential models the physical electron-molecule
scattering potential of CF$_4$ for elastic scattering, although electron correlation 
and target response effects are neglected.
The differences between the cross sections calculated using the overset-grid Kohn method with $\ell_{\mathrm{max}}=3$
in the subgrids and with $\ell_{\mathrm{max}}=$15, 25, and 35 in the
central grid are small, especially outside the region of the
resonance and except at very low energies.
In the region of the resonances (one $T_2$ resonance and about 10.0\,eV
and one $A_1$ resonance at about 11.5\,eV), the cross section calculated
at $\ell=15$ is slightly low. 
However, the shape of the resonance features is close to the converged
shape even with the lowest number of partial waves shown, which is not the case for the unconverged single-center expansion
result from the Schwinger calculations.

At energies below 10\,eV in Fig.~\ref{fig:CF4StaticExchange}, the
single-center expansion cross sections converge very slowly. 
At lower energies, the effects of of the potentials near the nuclei are
becoming increasingly important, as is evidenced by the fact that this 
is the most difficult region of the cross section  to converge in the
single-center expansion.
In Table~\ref{table:CF4SEat6eV}, we choose $E=6$\,eV to explore the
convergence of the single-center expansion Schwinger method, and compare
it to the convergence of the overset grid complex Kohn method.  Although no
exact result is available, the rates of convergence of the two methods can be judged 
at least approximately 
from the changes as $\ell_{\textrm{max}}$ is incremented that are shown in the table.  
By that measure, the total cross section from the overset grid Kohn method is converged to 0.01\,\AA$^2$
by $\ell=35$ in the central grid, and increasing the subgrid angular grid
from $\ell=3$ to $\ell=6$ has no effect.
The $A_1$ and $T_2$ components of the cross section converge to 0.0017
and 0.0030\,\AA$^2$ respectively.
In contrast, the single-center expansion results converge to only with 0.02\,\AA$^2$ 
at $\ell=200$.
To demonstrate conclusively that these methods are converging
to the same result to two digits beyond the decimal would require Schwinger calculations
calculations substantially beyond $\ell=200$  which rapidly become computationally prohibitive.

\begin{table*}
\caption{ \label{table:CF4SEat6eV}
The convergence of the integral cross section and its dominant symmetry 
components at $E=6$\,eV for CF$_4$ in the static-exchange approximation (in units of~\AA$^2$, with differences 
in parentheses). Schwinger variational cross
sections using a single center expansion are given from $\ell=60$ to $\ell=200$.  Overset-grid complex Kohn cross sections are
given for central grid $\ell=15$ to $\ell=45$, by which an effectively converged result
is obtained. }
\begin{ruledtabular}
\squeezetable
\begin{tabular}{rdddddddd}
\multicolumn{1}{c}{}&\multicolumn{8}{c}{Schwinger variational method, single-center expansion}\\
\cline{2-9}\\
             $\ell_{\mathrm{max}}$ &       A_1   &               &                E  &              &              T_2  &              &\mathrm{Total}&     \\
\hline
                              60   &     7.9166  &               &          0.2322   &              &          4.2152   &              &    21.0662   &             \\
                              80   &     7.8290  &   (-0.0876)   &          0.2315   &  (-0.0007)   &          4.0662   &  (-0.1491)   &    20.5299   &  (-0.5363)  \\
                             100   &     7.7752  &   (-0.0538)   &          0.2313   &  (-0.0002)   &          3.9974   &  (-0.0688)   &    20.2694   &  (-0.2606)  \\
                             120   &     7.7442  &   (-0.0309)   &          0.2313   &  (-0.0001)   &          3.9623   &  (-0.0351)   &    20.1330   &  (-0.1364)  \\
                             140   &     7.7259  &   (-0.0184)   &          0.2312   &  (-0.0000)   &          3.9426   &  (-0.0197)   &    20.0556   &  (-0.0774)  \\
                             160   &     7.7144  &   (-0.0115)   &          0.2312   &  (-0.0001)   &          3.9306   &  (-0.0120)   &    20.0081   &  (-0.0475)  \\
                             180   &     7.7075  &   (-0.0069)   &          0.2312   &   (0.0001)   &          3.9238   &  (-0.0068)   &    19.9807   &  (-0.0274)  \\
                             200   &     7.7027  &   (-0.0048)   &          0.2312   &  (-0.0000)   &          3.9189   &  (-0.0048)   &    19.9613   &  (-0.0193)  \\
\hline                                                                                                                                              
\multicolumn{1}{c}{}&\multicolumn{8}{c}{Complex Kohn variational method, overset-grid expansion}\\
\cline{2-9}\\
 $\ell_{\mathrm{max}}$ (cg/sg)   &       A_1   &               &                E  &              &              T_2  &              &\mathrm{Total}&     \\
\hline
                            15/3   &     7.7536  &               &          0.2107   &              &          3.8348   &              &    19.7149   &             \\
                            25/3   &     7.6605  &   (-0.0931)   &          0.2309   &   (0.0202)   &          3.8795   &   (0.0448)   &    19.8001   &   (0.0853)  \\
                            35/3   &     7.6907  &    (0.0302)   &          0.2314   &   (0.0005)   &          3.9083   &   (0.0288)   &    19.9179   &   (0.1178)  \\
                            35/6   &     7.6907  &   (-0.0000)   &          0.2314   &  (-0.0000)   &          3.9083   &  (-0.0000)   &    19.9178   &  (-0.0000)  \\
                            45/3   &     7.6890  &   (-0.0017)   &          0.2312   &  (-0.0002)   &          3.9053   &  (-0.0030)   &    19.9066   &  (-0.0112)  \\
\end{tabular}
\end{ruledtabular}
\end{table*}

\begin{table*}
\caption{ \label{table:CF4SEInterestingPoints}
Convergence of integral cross sections for the static-exchange 
potential of CF$_4$ using the Schwinger method with single center expansion 
and the overset-grid complex Kohn method for four points of interest: the $A_1$ components at 
0.01\,eV and 0.1\,eV, the $A_1$ resonance at 11.5\,eV, and the $T_2$ 
resonance at 10.0\,eV. 
}
\begin{ruledtabular}
\begin{tabular}{rdddddddd}
\multicolumn{1}{c}{}&\multicolumn{6}{c}{$A_1$}&\multicolumn{2}{c}{$T_2$}\\
\cline{2-7}\cline{8-9}\\
             $\ell_{\mathrm{max}}$ & 0.01\,\mathrm{eV} &                   & 0.1\,\mathrm{eV} &                   & 11.5\,\mathrm{eV} &                    & 10.0\,\mathrm{eV} &              \\
\hline
\multicolumn{1}{c}{}&\multicolumn{8}{c}{Schwinger variational method, single-center expansion}\\
                           60      &      25.1782     &                    &      24.6777    &                    &        6.8941    &                     &       6.3549     &               \\
                          120      &      22.8114     &     (-2.3667)      &      22.3826    &      (-2.2951)     &        7.8005    &       (0.9064)      &       6.1151     &     (-0.2398) \\
                          140      &      22.6240     &     (-0.1874)      &      22.2007    &      (-0.1819)     &        7.7918    &      (-0.0087)      &       6.1006     &     (-0.0145) \\
                          160      &      22.5088     &     (-0.1152)      &      22.0889    &      (-0.1118)     &        7.7837    &      (-0.0081)      &       6.0924     &     (-0.0082) \\
                          180      &      22.4438     &     (-0.0650)      &      22.0256    &      (-0.0633)     &        7.7766    &      (-0.0071)      &       6.0871     &     (-0.0053) \\
                          200      &      22.3975     &     (-0.0463)      &      21.9807    &      (-0.0450)     &        7.7718    &      (-0.0048)      &       6.0838     &     (-0.0034) \\
\hline                                                                                                                                              
\multicolumn{1}{c}{}&\multicolumn{8}{c}{Complex Kohn variational method, overset-grid expansion}\\
                         15/3      &      22.8492     &                    &      22.4204    &                    &        7.8700    &                     &       6.0061     &               \\
                         25/3      &      22.0215     &     (-0.8277)      &      21.6151    &      (-0.8053)     &        7.7112    &      (-0.1589)      &       6.0572     &      (0.0511) \\
                         35/3      &      22.2841     &      (0.2626)      &      21.8705    &       (0.2553)     &        7.7574    &       (0.0462)      &       6.0760     &      (0.0188) \\
\end{tabular}
\end{ruledtabular}
\end{table*}

\begin{figure}
\includegraphics[width=0.95\linewidth]{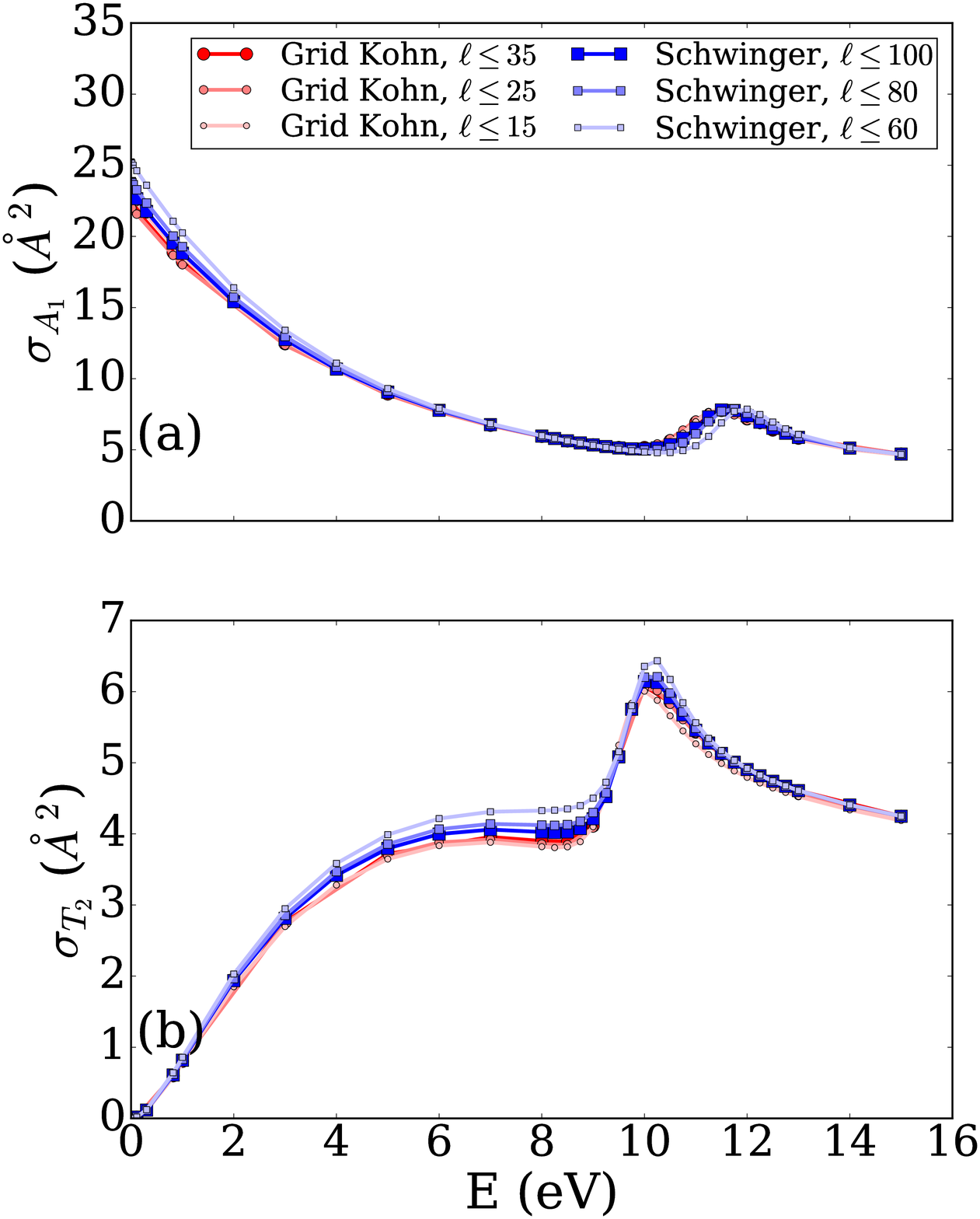}
\caption{(Color online)
The (a) $A_1$ and (b) $T_2$ components of the
static-exchange cross section of CF$_4$.
The grid Kohn method cross sections (red circles, various fading) and
single-center expanded Schwinger method cross sections (blue squares,
various fading) are compared.
\label{fig:CF4SEComponents}}
\end{figure}

We have also examined the symmetry components of the cross sections as a
function of energy, shown in Fig.~\ref{fig:CF4SEComponents}.
Here, we can see in more detail what is happening as the resonance
regions converge.
For the single-center expanded Schwinger method, the $A_1$ resonance is
moving to lower energies as it converges~\cite{lucchese1982studies}.
The overset-grid Kohn method, in comparison, converges its prediction of
the $A_1$ resonance energy closely by $\ell=15$.
At very low energies, the Schwinger method also converges slowly
compared to the grid Kohn method.
The behavior of the methods at the $T_2$ resonance is similar, with the
single-center expanded Schwinger method converging the magnitude of the
resonance slowly from $\ell=60$ to $\ell=100$, and the grid Kohn method
converging much more rapidly.
For energies below 10\,eV, the Schwinger $T_2$ cross sections also
converge more slowly compared to the grid Kohn cross sections.

\begin{figure}
\includegraphics[width=0.95\linewidth]{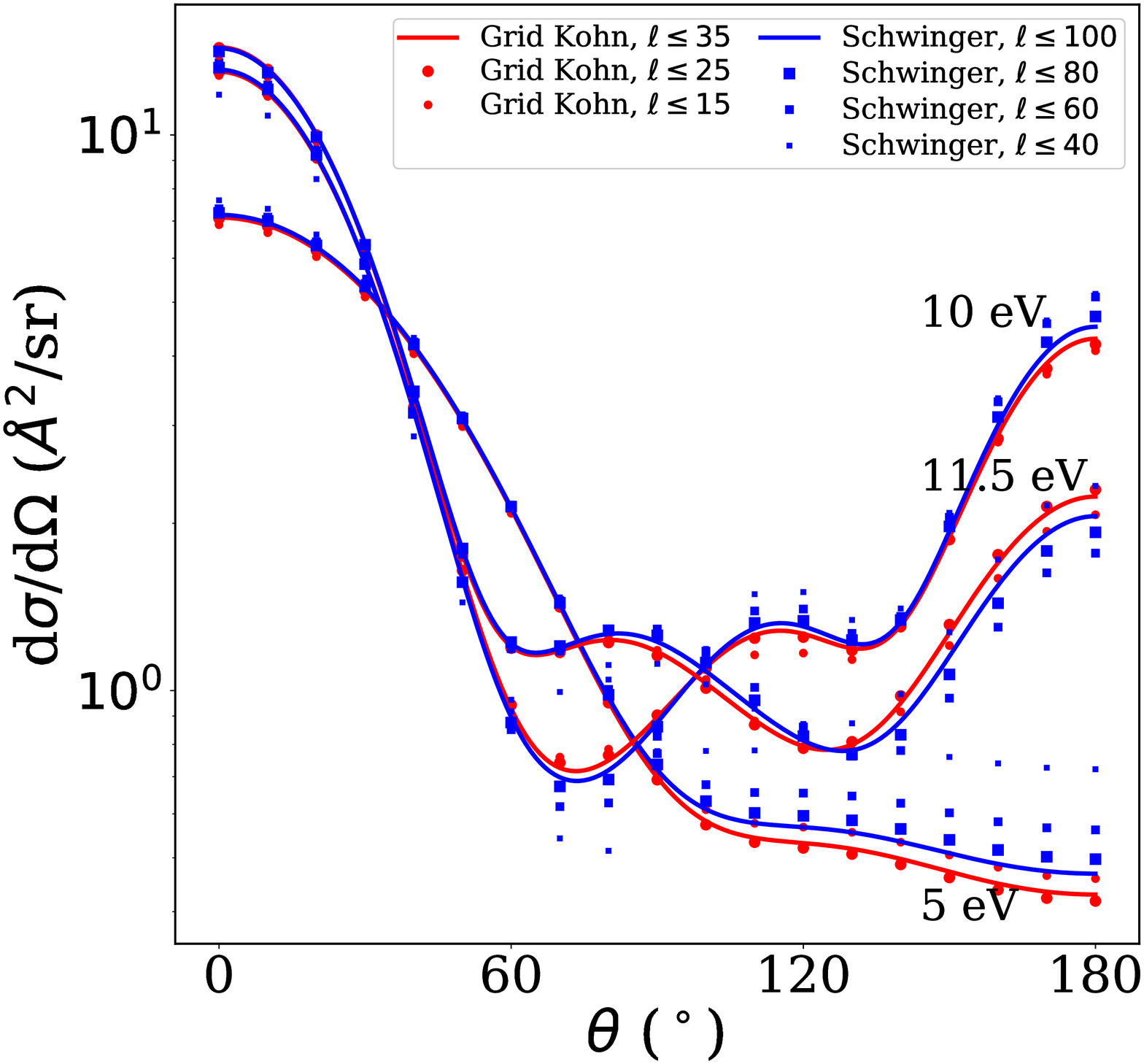}
\caption{(Color online)
Convergence of differential cross sections for CF$_4$ in the static-exchange approximation at 5 eV and at the energies
of the centers of the resonance features.  Complex Kohn results: red dots and solid red line with the converged results 
at $\ell_{max} = 35$. Schwinger with single-center expansion: blue dots with solid blue line at $\ell_{max} = 100$.
\label{fig:CF4SEDifferential}}
\end{figure}

The convergence of the differential cross section is shown in Fig.~\ref{fig:CF4SEDifferential}, both at 
the resonance features and at 5 eV.  As might be expected, the cross sections converge most slowly 
for back scattering and at the oscillations at intermediate angles for the resonance energies.  In all cases 
the convergence of the grid-based Complex Kohn with added partial waves on the master grid is dramatically faster
than the single-center expansion results.

In Table~\ref{table:CF4SEInterestingPoints}, we examine the convergence
of the single-center and overset-grid expansions for the following
points of interest: the $A_1$ components of the cross section at the
very low energies (i) 0.01 and (ii) 0.1\,eV, for which the potential near the 
nuclei must be well-resolved, (iii) the center of the $A_1$ resonance feature at 11.5\,eV, 
and (iv) the  center of the $T_2$ resonance at 10.0\,eV.
For the $A_1$ and $T_2$ resonance cross sections, the Schwinger method 
converges to the values 7.8\,\AA$^2$ and 6.1\,\AA$^2$ by $\ell=120$, respectively, 
and converges to 7.77\,\AA$^2$ and 6.08\,\AA$^2$, respectively,
with a tolerance of 0.005\,\AA$^2$ by $\ell=200$.
At $\ell=35$ for the central grid and $\ell=3$ for the subgrid, the
overset grid Kohn method predicts cross sections for these resonances of
7.76\,\AA$^2$ and 6.08\,\AA$^2$, in near perfect agreement with the
Schwinger results but with many fewer partial waves.
All of these comparisons are done with the maximum value of the radial variable, $r$,
 on the grids in both equal to 5.3~\AA, chosen purely for convenience
 to be just beyond the range of the exchange potential.  
Nonetheless, at at the lowest energies the cross sections from the Schwinger and Kohn
calculations agree only to within 0.1\,\AA$^2$ , suggesting that the remaining small differences
 between them are due to slower convergence with respect to partial waves at low energies
 of the A$_1$ component of the $T$-matrix seen in the table.

\section{Summary}
\label{sec:conclusion}

We have formulated an overset-grid version of the Complex Kohn variational approach to electron-polyatomic molecule 
collisions and demonstrated that it is a robust method for solving the scattering problem.  The key components  developed here,
all necessary for this formulation, are:  
\begin{itemize}
\item{An over-complete underlying spectral basis defined on overlapping subgrids and master grid.}
\item{Simultaneous expansion of the scattering trial wave function in spherical 
harmonics about multiple centers.}
\item{Expansion of the trial function in the basis $\phi^+_k = \left(\hat{G}^+_0 \hat{V}\right)^k \phi_0$ in 
which all the basis functions satisfy outgoing wave boundary conditions.}
\item{Accurate  operation of the free-particle Green's function and exchange operators 
based on Gauss-Lobatto quadratures adapted for integration over subintervals required by the slope discontinuity
in the radial Green's function.}
\end{itemize}
This grid-based method will allow the application of the Kohn variational approach both to larger systems and higher energies
and also removes the ``separable exchange'' approximation made in the previous implementation of the Complex Kohn method
for electron-molecule scattering.  Importantly, the over completeness of the underlying spectral basis from the union of the subgrids and
master grid produces no noticeable numerical pathologies. 

While the principal motivation to represent the trial wave function of the Kohn approach on overset grids 
is to gain the accuracy and generality of grid representation in this variational approximation to the 
scattering amplitudes, the basis we have chosen in order to apply outgoing-wave boundary conditions produces a 
formal property for this approach not previously exploited in any implementation of Kohn's original idea.   In the 
present method a Pad\'e approximant to the full solution of the scattering problem is being 
automatically constructed.  That property allows the convergence of the $T$-matrix with a basis 
of 20-30 functions defined by $(\hat{G}^+_0 \hat{V})^k \phi_0$ even for the largest grids used in these calculations.
 
This aspect of the method will be critical to the extension of this approach to multichannel calculations.   All of the
fundamental components
necessary for implementing the close-coupling version of the Complex Kohn method~\cite{mccurdy1989collisions} 
have been demonstrated here.   That extension to multiple scattering channels and the application of Coulomb 
boundary conditions within this framework are underway.

\begin{acknowledgments}
Work at the University of California Davis was supported by the U. S. Army Research Laboratory and the U. S. Army Research Office under grant number W911NF-14-1-0383.  Calculations presented here made use of the
resources of the National Energy Research Scientific Computing Center, 
a DOE Office of Science User Facility.
Computational resources were also provided by Texas A\&M High 
Performance Research Computing.
Work performed at Texas A\&M University was supported by the US Department of 
Energy Office of Basic Energy Sciences, Division of Chemical Sciences Contract 
DE-SC0012198.
\end{acknowledgments}

%

%
%
\end{document}